\documentclass{elsart}
\usepackage{graphicx}
\usepackage{amssymb}
\usepackage{bm}
\usepackage{amsmath}
\usepackage{enumerate}

\newcommand{\epl}{Europhys. Lett.\ }

\newcommand{\pra}{Phys. Rev. A\ }
\newcommand{\prb}{Phys. Rev. B\ }
\newcommand{\prl}{Phys. Rev. Lett.\ }

\newcommand{\oc}{Opt. Commun.\ }
\newcommand{\pr}{Phys. Rev.\ }

\newcommand{\jpb}{J. Phys. B\ }
\newcommand{\nat}{Nature\ }
\newcommand{\ap}{Adv. Phys.\ }
\newcommand{\jmo}{J. Mod. Opt.\ }

\newcommand{\bra}[1]{\langle#1|}
\newcommand{\ket}[1]{|#1\rangle}

\begin{document}

\begin{frontmatter}

\title{Numerical representation of quantum states in the positive-P and Wigner representations}

\author{M.~K. Olsen} 

\address{ARC Centre of Excellence for Quantum-Atom Optics, 
School of Physical Sciences, University of Queensland, Brisbane, 
QLD 4072, Australia} 

\author{A.~S. Bradley}

\address{Jack Dodd Centre for Quantum Technology, Department of Physics, University of Otago, PO Box 56, Dunedin, New Zealand }

\begin{abstract}

Numerical stochastic integration is a powerful tool for the investigation of quantum dynamics in interacting many body systems. As with all numerical integration of differential equations, the initial conditions of the system being investigated must be specified.
With application to quantum optics in mind, we show how various commonly considered quantum states can be numerically simulated by the use of widely available Gaussian and uniform random number generators. We note that the same methods can also be applied to computational studies of Bose-Einstein condensates, and give some examples of how this can be done.
 
\end{abstract}


\end{frontmatter}

\section{Introduction}
\label{sec:intro}

The theoretical study of non-equilibrium quantum many-body dynamics is a growing area, especially since the experimental achievement of trapped Bose-Einstein condensates. Many of the methods used for theoretically investigating condensates have been adapted from theoretical quantum optics~\cite{Danbook}, with varying degrees of success. One particular approximation technique that proved extremely successful in quantum optics is linearisation of the fluctuations about solutions of the classical equations of motion. This technique, if used appropriately, is a very powerful tool for the calculation of the steady-state spectra of intracavity parametric processes~\cite{Matthew}. However, in a dynamically evolving system, or one operating near phase transitions or critical points, this method can give incorrect answers~\cite{OPO,revive}. The validity of the approximation depends on three conditions. The first of these is that the solution of the classical equations is the same as the mean-field solution of the full quantum equations. The second and third are that the fluctuations about these solutions are in some sense small and that they can be represented as Gaussian, so that moments of higher than second order vanish. In the study of trapped Bose-Einstein condensates, the Hartree-Fock-Bogoliubov (HFB) method is a closely related approximation~\cite{HFB}, and therefore needs to be used with the same care as the linearised fluctuation approximation in quantum optics. 

When these conditions are not met, there are still a number of ways to proceed. In some very rare cases it may be possible to solve directly either a master equation for the density matrix, or even the Heisenberg equations of motion for the actual system operators. However, the most interesting quantum dynamics are not generally restricted to such cases. One set of methods which has been very successful is the phase-space representations originally used to develop stochastic differential equations in quantum optics~\cite{Qnoise}. These allow common classes of quantum Hamiltonians to be mapped via master and Fokker-Planck equations onto stochastic differential equations. In some cases the Fokker-Planck equation may be solved directly for a pseudoprobability distribution which then allows for the calculation of operator moments~\cite{Danbook,FPEsol}. Once again, these cases are rare and can often only be solved in the steady-state regime. The method of choice if we wish to obtain dynamical quantum information is then to numerically integrate the stochastic equations of motion. As with any numerical analysis of differential equations, this then requires that the initial conditions be specified, as these can have marked effects on the subsequent dynamics, in both optical~\cite{earlyg2,ocstate} and interacting atomic and molecular systems~\cite{RCstate,expandstate,LPHYS03,FockBEC,Bradley08a,Weiler08a,Wright08a}. In what follows we will begin with a brief outline of the theory behind the phase-space representations and then show how to numerically simulate some of the more common and useful initial quantum states of optics and condensed atom physics, in both the positive-P~\cite{P+} and Wigner representations~\cite{Wigner}.

\section{Phase-space representations of the density matrix}
\label{sec:fase}

Phase space techniques are a powerful tool to investigate the full quantum dynamics of interacting quantum systems in cases where it is impractical to solve either the Heisenberg equations of motion or the master (von Neumann) equation for the density matrix. Instead of working with operators or density matrices, they allow us to work directly with classical c-number variables, which are amenable to manipulation on available computers. Perhaps more importantly, the complexity of the computation scales with the number of interacting modes rather than with the size of the Hilbert space, which is often completely intractable. In fact, a single-mode quantum calculation has been performed using these methods for the order of $10^{23}$ interacting quanta~\cite{enorme}, which would be completely out of the question using other methods. There are a number of phase-space representations, among them being the Wigner representation~\cite{Wigner}, the Glauber-Sudarshan P representation~\cite{Roy,Sud}, the Q representation~\cite{Husimi} (sometimes known as the Husimi representation), the complex P representation~\cite{P+}, and the R representation~\cite{Roy}. The most useful for numerical work are the positive-P and truncated Wigner representations~\cite{Robert}, the latter being an approximation to the full Wigner representation.

\subsection{Truncated Wigner equations}
\label{subsec:Wminus}

Historically, the first of these phase-space representations was the Wigner representation~\cite{Wigner}, which was formulated as a pseudoprobability function for the position and momentum of a particle. Mathematically, the quadrature phase amplitudes of quantum optics are completely equivalent to position and momentum, so that the Wigner function is a frequently used tool for describing nonclassical states of bosonic fields. Quantum mechanical expectation values for operator products expressed in symmetrical order are found naturally in the Wigner representation as classical averages of the corresponding Wigner variables. As an example, making the correspondence between the single-mode annihilation operator $\hat{a}$ and the complex Wigner variable $\alpha$, we find that
\begin{equation}
\overline{\alpha^{\ast}\alpha} = \frac{1}{2}\langle \hat{a}^{\dag}\hat{a}+\hat{a}\hat{a}^{\dag}\rangle = N+\frac{1}{2},
\label{eq:avWig}
\end{equation}
where $N$ is the number of quanta in the mode. 
Given a general Hamiltonian which is some combination of bosonic creation and annihilation operators, ${\it H}$, we find the von Neumann equation as
\begin{equation}
i\hbar\frac{d\rho}{dt} = \left[{\it H},\rho\right],
\label{eq:vN}
\end{equation}
from which the equation of motion for the Wigner function, $W$, is found using the correspondence rules,
\begin{eqnarray}
\hat{a}\rho  \leftrightarrow \left(\alpha+\frac{1}{2}\frac{\partial}{\partial\alpha^{\ast}}\right)W,&\hspace{1cm}&
\hat{a}^{\dag}\rho  \leftrightarrow  \left(\alpha^{\ast}-\frac{1}{2}\frac{\partial}{\partial\alpha}\right)W,\nonumber\\
\rho\hat{a}  \leftrightarrow  \left(\alpha-\frac{1}{2}\frac{\partial}{\partial\alpha^{\ast}}\right)W,& &
\rho\hat{a}^{\dag}  \leftrightarrow  \left(\alpha^{\ast}+\frac{1}{2}\frac{\partial}{\partial\alpha}\right)W.
\label{eq:Wigcorrespond}
\end{eqnarray}
Following the standard methods~\cite{stochmeth}, as long as the equation found by the above procedure has no derivatives of higher than second order, it can be mapped onto a set of stochastic differential equations for the variables $\alpha$ and $\alpha^{\ast}$. Unfortunately, all interesting problems result in derivatives of third order or more and, although methods exist for mapping the resulting generalised Fokker-Planck equations onto stochastic difference equations~\cite{ourEPL}, these are not very useful in practice. A common practice is to truncate the partial differential equation for the Wigner function at second order, often justified by claiming that the effect of these terms is small. This procedure may be formally justified by requiring the system modes to be highly occupied, and results in stochastic differential equations in what is known as the truncated Wigner representation. If there are no second order derivatives, the resulting equations are regular and quantum noise enters via the initial Wigner distribution for the variables. In optical problems, this then becomes functionally equivalent to stochastic electrodynamics~\cite{trevor} and has been shown to give misleading results in some cases~\cite{Kinsler,asc}. This approximate method has also been used with some success in the study of Bose-Einstein condensates~\cite{Steel,Alice1,Alice2,Alice3} and is closely related to ``classical field methods", including the stochastic Gross-Pitaevski equation~\cite{SGPE1,SGPE2,Polkovnikov04a,cfields}. The appropriate initial states to use in the truncated Wigner equations are exactly the same as those that would be used in a full Wigner representation, with the approximations entering into the equations of motion.

\subsection{Positive-P representation}
\label{subsec:Pplus}

The Glauber-Sudarshan P representation~\cite{Roy,Sud} is another representation of the density matrix in terms of coherent states and gives averages of the phase-space variables which are equivalent to normally-ordered operator expectation values,
\begin{equation}
\overline{(\alpha^{\ast})^{m}\alpha^{n}} = \langle (\hat{a}^{\dag})^{m}\hat{a}^{n}\rangle.
\label{eq:Paverages}
\end{equation}
As photodetectors naturally measure normally-ordered averages, this would at first glance seem to be an extremely useful representation. It does, however, have two serious drawbacks. The first is that it is difficult to represent any state which is ``more quantum" than a coherent state, as these do not possess positive and analytic P-functions. Although a P-function can be written for any quantum state in terms of generalised functions~\cite{Klauder}, it is difficult to see how to sample these numerically. The second drawback arises when we consider the P-representation Fokker-Planck equation, found using the operator correspondences
\begin{eqnarray}
& &\hat{a}\rho  \leftrightarrow  \alpha P,\hspace{1cm}
\hat{a}^{\dag}\rho  \leftrightarrow  \left(\alpha^{\ast}-\frac{\partial}{\partial\alpha}\right) P,\nonumber\\
& &\rho\hat{a} \leftrightarrow \left(\alpha-\frac{\partial}{\partial\alpha^{\ast}}\right) P,\hspace{1cm}
\rho\hat{a}^{\dag} \leftrightarrow  \alpha^{\ast}P.
\label{eq:Pcorrespondences}
\end{eqnarray}
It is readily seen that, for any interesting problem, the resulting Fokker-Planck equation will not have a positive-definite diffusion matrix and therefore will not be able to be mapped onto stochastic differential equations. The positive-P representation~\cite{P+} was developed to circumvent this problem by using a doubled phase space. For Hamiltonians which lead to derivatives of no higher than second order, this results in a Fokker-Planck equation which always has a positive-definite diffusion matrix and therefore can always be mapped onto stochastic differential equations. The price which has to be paid is that, instead of having $\alpha$ and $\alpha^{\ast}$ as complex conjugate variables, the variables corresponding to this pair become independent. These are written in various ways, but in this article we will write the pair as $\alpha$ and $\alpha^{+}$, and the appropriate equations can be found by naively using the P representation correspondences of Eq.~\ref{eq:Pcorrespondences} and then substituting $\alpha^{+}$ for $\alpha^{\ast}$. The independence of the variables can cause serious stability problems with the numerical integration, but for problems where the integration converges, the positive-P representation is an extremely powerful theoretical tool~\cite{Deuar07a}. As a final remark, we note that a method has been developed for mapping Hamiltonians which would give higher than second order derivatives in a generalised Fokker-Planck equation onto stochastic difference equations~\cite{P++}, which is useful for analysing processes such as third harmonic generation~\cite{H3} and others which go beyond the common three and four-wave mixing processes of quantum optics and trapped ultra-cold gases.

\section{Representation of quantum states}
\label{sec:initialise}

As always with the numerical integration of differential equations, initial conditions must be specified. In this section we will describe how this can be done for a number of quantum states which typically arise in quantum optics and BEC problems. We note here that we do not specify the type of stochastic differential equation, and all our results may be used as initial conditions for either regular (common in the truncated Wigner for systems with quadratic Hamiltonians) or stochastic (common in positive-P and truncated Wigner for damped systems) differential equations. For simplicity, we will consider single-mode representations of the states, with a multi-mode extension being relatively straightforward. We will also show how our choices for these states represent the correct quantum statistics, in terms of both quadrature and intensity variances. Note that we use quadrature definitions such that
\begin{equation}
\hat{X} = \hat{a}+\hat{a}^{\dag},\hspace{1cm}\hat{Y}=-i\left(\hat{a}-\hat{a}^{\dag}\right),
\label{eq:quadratures}
\end{equation}
so that the coherent state variances are equal to $1$. 

\subsection{Coherent states}
\label{sec:coherent}

The simplest initial condition to model in the positive-P representation is a coherent state, $|\beta\rangle$, defined by $\hat{a}|\beta\rangle=\beta|\beta\rangle$. When the Glauber P-function for a given state is well-behaved, we may also use this as the positive-P function. This means that a coherent state can be represented by the pseudoprobability distribution $P(\alpha,\alpha^{+})=\delta(\alpha-\beta)\delta(\alpha^{+}-\beta^{\ast})$. Numerically, this means using the same complex conjugate pair, $\beta$ and $\beta^{\ast}$, for each of the stochastic trajectories. This state is an appropriate choice to represent a well-stabilised laser output and is also commonly used as an initial condition for single-mode BEC analyses.

The Wigner function for the coherent state is a Gaussian,
\begin{equation}
W(\alpha,\alpha^{\ast}) = \frac{2}{\pi}\exp\left(-2|\alpha-\beta|^{2}\right).
\label{eq:coWig}
\end{equation}
This is simply represented numerically by choosing the initial condition for each trajectory as 
\begin{equation}
\alpha=\beta+\frac{1}{2}\left(\eta_{1}+i\eta_{2}\right),
\label{eq:conumW}
\end{equation}
where the $\eta_{j}$ are sampled from a real normal Gaussian distribution, such as is given by the Matlab function randn. These have the correlations $\overline{\eta_{j}}=0$ and $\overline{\eta_{j}\eta_{k}}=\delta_{jk}$, where the overline denotes an average over many samples. It is readily shown that $\overline{\alpha}=\beta$ and $\overline{|\alpha|^{2}}=|\beta|^{2}+\frac{1}{2}$ and that the quadrature variances both give $1$, as required.

\subsection{Thermal or chaotic states}
\label{sec:thermal}

These states, which can be used to represent, for example, a mechanical oscillator in a thermal bath~\cite{pendcav}, have a particularly simple P-function, with
\begin{equation}
P(\beta)=\frac{1}{\pi\overline{n}}\exp(-|\beta|^{2}/\overline{n}),
\label{eq:Pchaos}
\end{equation}
where $\overline{n}$ is the average number present in the mode~\cite{Danbook}. We see immediately that, as expected, there is no phase information in this state. The appropriate distribution can be sampled from a normal Gaussian distribution multiplied by $\sqrt{\overline{n}}$ and a random phase term,
\begin{equation}
\alpha = \sqrt{\overline{n}}\,\eta\times\exp\left(2\pi i\zeta\right),
\label{eq:Pthermal}
\end{equation}
where $\eta$ is again a normal Gaussian variable and $\zeta$ is uniformally distributed on $[0,1]$. Gaussian and uniform variables may be easily sampled using, for example, the Matlab functions {\tt randn} and {\tt rand} respectively. Numerical checks of this distribution show that, with sufficient samples $(\gtrsim 10^{4})$, it reproduces well the intensity, $\overline{n}$, and the quadrature variances, $V(\hat{X})=V(\hat{Y})=2\overline{n}+1$.
If the state is a mixture of coherent and chaotic states, i.e. a chaotic state 
with a coherent displacement, the P-function is written as
\begin{equation}
P(\beta)=\frac{1}{\pi\overline{n}}\exp(-|\beta-\beta_{0}|^{2}/\overline{n}),
\label{eq:mix}
\end{equation}
where $\beta_{0}$ is the coherent displacement. This may be easily sampled as
\begin{equation}
\alpha = \beta_{0}+\sqrt{\overline{n}}\,\eta\times\exp\left(2\pi i\zeta\right),
\label{eq:Pthermaldisplace}
\end{equation}
where the random variables are as in Eq.~\ref{eq:Pthermal}.

The Wigner function for the chaotic state may be found as a convolution of the P-function with a Gaussian of standard deviation one-half. In this case, where the P-function is itself a Gaussian, this results in a broader Gaussian, which can be sampled via
\begin{equation}
\alpha = \sqrt{\overline{n}+1/2}\,\eta\times\exp\left(2\pi i\zeta\right),
\label{eq:Wtherm}
\end{equation}
where the random terms are the same as in Eq.~\ref{eq:Pthermal}. Samples from this distribution reproduce well the required intensities and quadrature variances, always remembering that the Wigner distribution represents symmetrically ordered operator products.

\subsection{Squeezed states}
\label{sec:squeezed}

Squeezed states are those in which the variance of one quadrature of the field is below the coherent state value. Due to the Heisenberg uncertainty principle, which requires that $V(\hat{X})V(\hat{Y})\geq 1$, this means that the variance in the conjugate quadrature must be greater than that of a coherent state~\cite{NatureDan}. They can now be routinely produced in the laboratory and are a useful resource for such things as nonclassical pumping of parametric processes, outcoupling squeezed atom laser beams~\cite{Simonsqueeze} and quantum-limited measurement~\cite{measure}. Theoretically, a minimum uncertainty squeezed state with $V(\hat{X})V(\hat{Y})= 1$ is defined by the action of the squeezing operator,
\begin{equation}
S(\epsilon)=\exp{[(\epsilon^*)^2a^2/2-\epsilon^2(a^\dag)^2/2]},
\label{eq:Sopdef}
\end{equation}
on the vacuum, followed by the coherent displacement operator,
\begin{equation}
D(\eta)S(\epsilon)|0\rangle = |\eta,\epsilon\rangle,
\label{eq:squeezeop}
\end{equation}
where $\epsilon=r\mbox{e}^{2i\phi}$~\cite{Danbook}. The role of the squeeze factor, $r$, is apparent when we look at the quadrature variances (setting $\phi=0$ for convenience),
\begin{equation}
V(\hat{X}) = \mbox{e}^{-2r},\hspace{1cm}V(\hat{Y}) = \mbox{e}^{2r}.
\label{eq:squeezevars}
\end{equation}
Another point to note is that the squeezing process adds quanta to the mode, so that 
\begin{equation}
\langle\eta,\epsilon|\hat{a}^{\dag}\hat{a}|\eta,\epsilon\rangle = |\eta|^{2}+\sinh^{2}r.
\label{eq:addsqueeze}
\end{equation} 

We will demonstrate here how to develop a numerical simulation of squeezed states using a canonical expression for an arbitrary positive-P function~\cite{P+}. 
Given a density matrix $\hat{\rho}$, a particular form of the positive-P function is
\begin{equation}
P(\alpha,\alpha^+)=\frac{1}{4\pi^2}|\bra{(\alpha+(\alpha^+)^*)/2}\hat{\rho}\ket{(\alpha+(\alpha^+)^*)/2}|^2\mbox{e}^{-|\alpha-(\alpha^+)^*|^2/4}\;.
\label{eq:Pcanonical}
\end{equation}
We now use the linear transformation
\begin{equation}
\mu = (\alpha+(\alpha^+)^*)/2, \hspace{1cm}
\gamma = (\alpha-(\alpha^+)^*)/2,
\label{eq:transform}
\end{equation}
which has the Jacobian
\begin{equation}
\left|\frac{\partial (\alpha_x,\alpha_y,\alpha^+_x,\alpha^+_y)}{\partial (\mu_x,\mu_y,\gamma_x,\gamma_y)}\right|=4,
\label{eq:jacobian}
\end{equation}
so that the normalised distribution in terms of the new variables becomes
\begin{equation}
P(\mu,\gamma)=\frac{|\langle\mu|\eta,\epsilon\rangle|^2}{\pi}\frac{e^{-|\gamma|^2}}{\pi}.
\label{eq:newdist}
\end{equation}
The displacement property
\begin{equation}
D^\dag(\mu)D(\eta)=D^\dag(\mu-\eta)\exp{[-i{\rm Im}(\mu\eta^*)]},
\label{eq:codisplace2}
\end{equation}
can then be used to obtain
\begin{equation}
\label{eq:matrixels}
\langle\mu|\eta,\epsilon\rangle=\langle 0|D^\dag(\mu-\eta)S(\epsilon)|0\rangle\exp{[-i{\rm Im}(\mu\eta^*)]}.
\end{equation}
It is now convenient to make another change of variables by setting
\begin{equation}
\mu-\eta = e^{i\phi}\nu,
\label{eq:anotherchange}
\end{equation}
and also make use of the normally ordered form of the squeeze operator, 
\begin{equation}
S(r,\phi)=(\cosh{r})^{-1/2}\exp{\left(-\frac{\Gamma}{2}a^{\dag2}\right)}\exp{[-{\rm ln}(\cosh{r})a^\dag a]}\exp{\left(\frac{\Gamma^*}{2}a^2\right)},
\label{eq:normalsqueeze}
\end{equation}
where $\Gamma=\mbox{e}^{2i\phi}\tanh{(r)}$. Making these substitutions in Eq.~\ref{eq:matrixels}, we find
\begin{equation}
|\langle\mu|\eta,\epsilon\rangle|^2=\frac{e^{-|\nu|^2-(\nu^2+\nu^{*2})(\tanh{r})/2}}{\cosh{(r)}},
\end{equation}
and finally arrive at the separable Gaussian form
\begin{equation}
\label{eq:Pfinal}
P(\nu,\gamma)=\frac{e^{-\nu_x^2/(e^{-r}\cosh{r})}}{\sqrt{\pi e^{-r}\cosh{r}}}\frac{e^{-\nu_y^2/(e^{r}\cosh{r})}}{\sqrt{\pi e^{r}\cosh{r}}}\frac{e^{-|\gamma|^2}}{\pi},
\end{equation}
which can be sampled using standard methods. We sample Eq.~\ref{eq:Pfinal} and invert to find the appropriate random variables for the positive-P distribution initial condition as
\begin{eqnarray}
\alpha&=&e^{i\phi}\nu+\gamma+\eta,\\
\alpha^+&=&e^{-i\phi}\nu^*-\gamma^*+\eta^*,
\end{eqnarray}
where, given Gaussian random variables satisfying $\overline{n_j}=0$ and $\overline{n_{j}n_{k}}=\delta_{jk}$, 
\begin{eqnarray}
\gamma&=&\frac{1}{\sqrt{2}}(n_1+in_2),\\
\nu&=&\sqrt{\frac{e^{-r}\cosh{r}}{2}}n_3+i\sqrt{\frac{e^r\cosh{r}}{2}}n_4.
\label{eq:finalmente}
\end{eqnarray}
This distribution can be checked numerically and gives the appropriate values for the quadrature variances and intensities, with more samples needed for accuracy as the squeezing parameter becomes larger.

The squeezed states are very simply modelled in the Wigner representation, by deforming the coherent state distribution in the appropriate manner. For a squeezed state with coherent dispalcement $\eta$, this is done by sampling
\begin{equation}
\alpha = \eta+\frac{1}{2}\left(\zeta_{1}\mbox{e}^{-r}+i\zeta_{2}\mbox{e}^{r}\right),
\label{eq:Wsqueeze}
\end{equation}
with the $\zeta$ being real normal Gaussian variables. A numerical check of this distribution shows that it also reproduces the analytically calculated intensities and quadrature variances very accurately.

\subsection{Number or Fock states}
\label{sec:Fock}

The Fock state is a quantum state with a fixed number of quanta. In optics, for example, the decay of an excited two-level atom can give a one photon Fock state. Here we will give a method which was previously used to model spontaneous emission from single-atom bosonic states~\cite{Stojan} in the positive-P representation. 
The Fock state with $n$ quanta has density operator $\hat{\rho}=|n\rangle\langle n|$, which we will again sample using Eq.~\ref{eq:Pcanonical}. Introducing the new variables
\begin{equation}
\mu=\frac{\alpha-(\alpha^{+})^{\ast}}{2}\:\:\mbox{and}\:\:\gamma=\frac{\alpha+(\alpha^{+})^{\ast}}{2},
\label{eq:newvars}
\end{equation}
we find the separable expression
\begin{eqnarray}
P(\mu,\gamma) &=& \frac{\mbox{e}^{-|\gamma|^{2}}}{\pi}\frac{|\mu|^{2n}\mbox{e}^{-|\mu|^{2}}}{\pi n!}\nonumber\\
&=& \frac{\mbox{e}^{-|\gamma|^{2}}}{\pi}\frac{\Gamma(|\mu|^{2},n+1)}{\pi},
\label{eq:gamma}
\end{eqnarray}
where 
\begin{equation}
\Gamma(x,n) = \frac{\mbox{e}^{-x}x^{n-1}}{(n-1)!}
\label{eq:gammadef}
\end{equation}
is the Gamma distribution. Once again $\gamma=(n_1+i n_2)/\sqrt{2}$ is easily sampled via standard methods, while the Gamma distribution may be easily and efficiently sampled using a method given by Marsaglia and Tsang~\cite{Marsaglia} to give $z=|\mu|^{2}$, and $\mu=\sqrt{z}\;\mbox{e}^{i\theta}$, where $\theta$ is uniform on $[0,2\pi)$. We then invert to find
\begin{equation}
\alpha = \mu+\gamma\:\:\mbox{and}\:\:\alpha^{+}=\mu^{\ast}-\gamma^{\ast},
\label{eq:inversion}
\end{equation}
which are now correctly distributed to represent the Fock state $|n\rangle$. Numerical checks once again show that the intensities and variances are represented well if enough samples are taken. As seen in Ref.~\cite{Stojan}, the use of this method for the initial state also leads to analytically known dynamics being reproduced.

To model a Fock state in the Wigner representation, we adapt an approximation developed by 
Gardiner {\em et al.\/}~\cite{SGPE1} which allows us to approximately represent these without having to deal with negative pseudoprobabilities.
The Wigner function for the Fock state $|N\rangle$ is
\begin{equation}
W(\alpha,\alpha^*)=2\frac{(-1)^N}{\pi}\exp(-2|\alpha|^2)L_N(4|\alpha|^2),
\label{eq:WigN}
\end{equation}
where $L_N$ is the Laguerre polynomial of order $N$. This distribution is oscillatory and can obviously be either positive or negative, so cannot be easily 
simulated numerically. However, in the large $N$ regime Gardiner has made the observation~\cite{SGPE1} that the cumulative distribution behaves very
like a step function centered at $|\alpha|^2=N$. This distribution
can then be approximated by a Gaussian which gives the right moments for the mean and variance and approximates the higher moments well. The appropriate distribution is
\begin{equation}
P_N(n,\theta)=\sqrt{\frac{2}{\pi}}\exp{\left(-\frac{(n-N-1/2)^2}{2(1/4)}\right)},
\label{eq:WigGauss}
\end{equation}
where we have taken $\alpha=\sqrt{n}e^{i\theta}$, with $\theta$
uniform on $[0,2\pi)$.
The first three moments of this distribution are
\begin{eqnarray}\label{eq:Gmom}
\overline{\alpha^{*}\alpha} &=& N+\frac{1}{2}\\
\overline{\alpha^{*2}\alpha^2} &=& \left(N+\frac{1}{2}\right)^2+\frac{1}{4}\\
\overline{\alpha^{*3}\alpha^3} &=& \left(N+\frac{1}{2}\right)^3+\frac{3}{4}\left(N+\frac{1}{2}\right),
\end{eqnarray}
so that mean and variance are in agreement with what we expect analytically. We
can now show that such an approximation in fact generates all
moments of (\ref{eq:WigN}), up to a correction of order $1/N^2$, which
is negligible for large $N$.

Using the differential recursion relation for the Laguerre polynomials,
\begin{equation}\label{eq:diff}
x\frac{d}{dx}L_N(x)=N(L_N(x)-L_{N-1}(x)),
\end{equation}
a recursion relation for arbitrary even moments can be found.
Writing
\begin{equation}
(\overline{\alpha^{*m}\alpha^m})_{\scriptscriptstyle
N}=2\frac{(-1)^N}{\pi}\int
d^2\alpha\;\exp{(-2|\alpha|^2)}L_N(4|\alpha|^2)|\alpha|^{2m},
\end{equation}
we can use (\ref{eq:diff}) to find
\begin{eqnarray}\label{eq:rec}
(\overline{\alpha^{*m}\alpha^m})_{\scriptscriptstyle
N} &=& \frac{N+m}{2}(\overline{\alpha^{*m-1}\alpha^{m-1}})_{\scriptscriptstyle
N}+\frac{N}{2}(\overline{\alpha^{*m-1}\alpha^{m-1}})_{\scriptscriptstyle
N-1}.
\end{eqnarray}
The first three moments are
\begin{eqnarray}\label{eq:Truemom}
(\overline{\alpha^{*}\alpha})_{\scriptscriptstyle N} &=& N+1/2,\\
(\overline{\alpha^{*2}\alpha^2})_{\scriptscriptstyle
N} &=& (N+1/2)^{2}+1/4,\\
(\overline{\alpha^{*3}\alpha^3})_{\scriptscriptstyle
N}&=&(N+1/2)^{3}+\frac{5}{4}(N+1/2),
\end{eqnarray}
and comparison with (\ref{eq:Gmom}) shows that the Gaussian approximation is exact
for the mean and variance, and accurate to $O(1/N^2)$ for $m=3$.
It is easily shown by induction on $m$ that the exact moments satisfy
\begin{equation}
(\overline{\alpha^{*m}\alpha^m})_{\scriptscriptstyle
N}=(N+1/2)^m+O(N^{m-2}),
\end{equation}
so that the correction is always
of order $1/N^2$ relative to the leading term.
Returning to the Gaussian approximation, we see that it gives
\begin{eqnarray}
\overline{\alpha^{*m}\alpha^m} &=& \sqrt{\frac{2}{\pi}}\int_{-\infty}^{\infty}dz\;(z+N+1/2)^m\;\exp{(-2z^2)}\nonumber\\
&=&\sqrt{\frac{2}{\pi}}\int_{-\infty}^{\infty}dz\;\left[(N+1/2)^m +
m(N+1/2)^{m-1}z + O(N^{m-2})\right]\exp{(-2z^2)}\nonumber\\
&=&(N+1/2)^m+O(N^{m-2}),
\end{eqnarray}
which will be an adequate
description of the number state statistics for large $N$.

To simulate this distribution numerically, consider the choice
\begin{equation}
\alpha=p+q\eta,
\label{eq:Wigrand}
\end{equation}
where $\eta$ is a normal Gaussian random variable, and $p$ and $q$ are yet to be determined. As we are using a Gaussian approximation, 
it is sufficient to reproduce the first two moments of $\alpha^{2}$. (Note that $\alpha$ is a real variable here, with the phase distribution 
to be added later). We need to reproduce $\overline{\alpha^{2}}=N+1/2$ and $\overline{\alpha^{4}}=(N+1/2)^{2}+1/4$.
The choices
\begin{equation}
p=\frac{1}{2}\left(2N+1+2\sqrt{N^{2}+N}\right)^{1/2},
\label{eq:pfN}
\end{equation}
and
\begin{equation}
q=\frac{1}{4p}.
\label{eq:qandp}
\end{equation} 
reproduce the required distribution to a high degree of accuracy.
The $\alpha$ thus chosen are then multiplied by the factor $\exp(2i\pi\xi)$, where $\xi$ is randomly chosen from the uniform distribution $[0,1)$. 

\subsection{Crescent states}
\label{sec:crescent}

The crescent state is given this name because its Wigner contours are sheared in phase-space due to a $\chi^{(3)}$ (Kerr) nonlinearity, and is consistent with quantum states which have been proposed for trapped Bose-Einstein condensates~\cite{Jacob,ssc,chernyak}, where s-wave scattering is equivalent to a Kerr nonlinearity. In this section we will treat the Wigner distribution first, because we will use the corresponding Q-distribution to sample a positive-P distribution for the crescent state.

In the Wigner representation a sheared state with coherent displacement $\alpha_{0}$ is simulated by beginning with the squeezed state representation given previously, and transforming this by a factor $\exp(iq\eta_{3})$, where $q$ is the shearing factor,
\begin{equation}
\alpha = \left[\alpha_{0}+\frac{1}{2}\left(\eta_{1}\mbox{e}^{-r}+i\eta\mbox{e}^{r}\right)\right]\mbox{e}^{iq\eta_{3}}.
\label{eq:Wigshear}
\end{equation}
The real noise terms have the correlations
\begin{equation}
\overline{\eta_{j}}=0,\:\:\overline{\eta_{i}\eta_{j}}=\delta_{ij}.
\label{eq:normalgauss}
\end{equation}
Numerical checks of distributions produced using these methods again show that they give the expected values for average numbers and quadrature variances.

The Q-distribution $Q(\mu,\mu^*)$ can be simulated as a simple broadening of the Wigner distribution, so that we sample
\begin{equation}
\mu = \left[\alpha_{0}+\frac{1}{\sqrt{2}}\left(\eta_{1}\mbox{e}^{-r}+i\eta\mbox{e}^{r}\right)\right]\mbox{e}^{iq\eta_{3}}.
\label{eq:Qshear}
\end{equation}
We can then make use of Eq. (\ref{eq:Pcanonical}) to construct samples of the corresponding positive-P distribution:
transforming to the variables
\begin{equation}
\mu = (\alpha+(\alpha^+)^*)/2, \hspace{1cm}
\gamma = (\alpha-(\alpha^+)^*)/2,
\label{eq:transform2}
\end{equation}
we have
\begin{equation}
P(\mu,\gamma)=\frac{\langle\mu|\rho|\mu\rangle}{\pi}\frac{e^{-|\gamma|^2}}{\pi}= Q(\mu,\mu^*)\frac{e^{-|\gamma|^2}}{\pi}.
\end{equation}
so that given the Q-function samples (\ref{eq:Qshear}), and $\gamma=(n_1+i n_2)/\sqrt{2}$, we have the crescent state sampling for the positive-P distribution
\begin{equation}
\alpha = \mu+\gamma\:\:\mbox{and}\:\:\alpha^{+}=\mu^{\ast}-\gamma^{\ast}.
\label{eq:inversion2}
\end{equation}
\subsection{Interacting many body states}
Generating the many body states appropriate for ultra-cold Bose gas simulations is difficult to illustrate within the single mode approach taken in this article. We note, however, that a number of Wigner states which are commonly used in modelling Bose-Einstein condensates may be sampled relatively easily and we provide a brief outline of currently available methods. We refer the reader to Ref.~\cite{cfields} for a detailed review of Wigner sampling methods for Bose gases. The methods may be summarized as follows
\begin{enumerate}[ 1) ]
\item {\em Coherent state}.---At zero temperature, a first approximation to the state of a BEC is a coherent state. In a continuous field theory the modes orthogonal to the condensate must necessarily contain vacuum noise in the Wigner representation. A simple and effective means to construct a coherent state is to simply add this noise, corresponding to half a quanta per mode, to the appropriate mean field solution of the GPE.
\item {\em Bogoliubov state}.---Steel \etal~\cite{Steel} demonstrated that an improved approximation at low temperatures, the Bogoliubov state, may be readily constructed from a stationary solution of the Gross-Pitaevskii equation once the Bogoliubov modes are known. This approach has also been extended to include $U(1)$ symmetry constraints imposed by number conservation by Sinatra \etal~\cite{Alice1,Alice2,Alice3}.
\item {\em Adiabatic mapping}.---Polkovnikov and Wang~\cite{Polkovnikov04a} used the quantum adiabatic theorem to show that interacting states at zero temperature may be obtained by sampling the appropriate noninteracting state and then adiabatically ramping up interactions to the desired final value.
\item {\em High temperature states}.---In the vicinity of $T_c$ a first approximation for the Wigner distribution is given by the non-interacting Bose-Einstein distribution. This may be used as a starting point for evolution according to the stochastic Gross-Pitaevskii equation~\cite{Bradley08a,Weiler08a,SGPE1,SGPE2} which evolves the Bose field toward a sample from the grand canonical ensemble for the many body system.
\end{enumerate} 

\section{Conclusions}
\label{sec:conclude}

In conclusion, we have shown how to take numerical samples in phase space of some of the most commonly appearing quantum states in quantum and atom optics. These techniques are important when we wish to investigate the effects of different initial states on dynamical quantum processes and possess straightforward generalisations to many-mode problems. These methods will become more useful and important as the dynamics and quantum features, such as entanglement, of interacting many-body systems are further investigated. 

\section*{Acknowledgments}
This research was supported by the Australian Research Council and the New Zealand Foundation for Research, Science and Technology.

\end{document}